\documentclass[%
 reprint,
nofootinbib,
 amsmath,amssymb,
 aps,
pra,
floatfix,
]{revtex4-2}
\usepackage{physics}
\usepackage{amsmath}
\usepackage{amsfonts} 
\usepackage{amssymb} 
\usepackage{float}
\usepackage{mathtools}
\usepackage{graphicx}
\usepackage{dcolumn}
\usepackage{bm}
\usepackage{hyperref}
\usepackage{svg}

\begin{document}

\preprint{APS/123-QED}

\title{Haar-random-state statistics as a guide to state preparation in globally controlled Rydberg atom arrays}

\author{Edison S. Carrera}
\email{ecarrera@ipht.fr}

\author{Grégoire Misguich}
\email{gregoire.misguich@ipht.fr}
 \affiliation{Université Paris-Saclay, CEA, CNRS, Institut de Physique Théorique, 91191 Gif-sur-Yvette, France}

\date{\today}

\begin{abstract}
Rydberg atom arrays are powerful platforms for studying quantum many-body systems. We consider the Rydberg-Ising Hamiltonian in a ring geometry and numerically study ensembles of states generated by random global pulse sequences subject to hardware constraints and fixed evolution times. We compare the statistical properties of such states with those of Haar-random states within the relevant lattice symmetry sector. In the strong-interaction regime (short interatomic distance), the dynamics is governed by an effective blockade that restricts Hilbert-space exploration and limits entanglement growth. In this regime, level-spacing statistics of the entanglement spectrum are close to random-matrix predictions, while the distribution of measurement probabilities deviates from Porter--Thomas behavior. For weaker interactions (larger interatomic distance), the system approaches Haar-like statistics at long times, as reflected in entanglement entropy, entanglement-spectrum statistics, and the distribution of measurement probabilities. At intermediate interactions, this behavior is observed on experimentally relevant timescales. Motivated by this observation, we investigate whether individual target states can be efficiently prepared using quantum optimal control in this regime. Employing target states drawn from an ensemble with a broad entropy distribution, we observe high fidelities (infidelities between \(10^{-5}\) and \(3\times10^{-2}\) for nine spins). Within this ensemble, the fidelity decreases with the entanglement entropy of the target state, indicating that highly entangled targets are more difficult to prepare under the considered finite-time control constraints.
\end{abstract}

\maketitle

\section{Introduction}
\label{sec:introduction}

Rydberg atom arrays have emerged as versatile platforms for the quantum simulation of correlated many-body systems~\cite{saffmanQuantumInformationRydberg2010,henrietQuantumComputingNeutral2020,Browaeys2020,MORGADO_QuantumSimulationComputing_2021}. Their combination of single-site resolution, tunable interactions, and scalability has enabled the realization of a wide range of phenomena, including many-body spin dynamics~\cite{labuhnTunableTwodimensionalArrays2016,bernienProbingManybodyDynamics2017,guardado-sanchezProbingQuenchDynamics2018,Bluvstein2021, Scholl2021,LECLERC_OnetooneQuantumSimulation_2026}, quantum phase transitions~\cite{bernienProbingManybodyDynamics2017,ebadiQuantumPhasesMatter2021, Keesling2019, Zhang2025,  Manovitz2025,LECLERC_OnetooneQuantumSimulation_2026}, spontaneous symmetry breaking~\cite{chenContinuousSymmetryBreaking2023}, topological phases~\cite{Kanungo2022, Semeghini2021, deLsleuc2019}, spin-squeezed states~\cite{Bornet2023} and resource states~\cite{Qin2025}.
These platforms have also been used to study interacting fermionic systems~\cite{MICHEL_HubbardPhysicsRydberg_2024,JULIA-FARRE_HybridQuantumclassicalAnalog_2025}.

A central ingredient in many of these applications is the ability to control the system's Hamiltonian. This enables the preparation of quantum states with desired properties and the exploration of specific dynamical regimes. Control protocols can rely on (quasi-)adiabatic evolution or on rapid variations of the Hamiltonian that drive the system out of equilibrium.

In quantum control theory, a central question is which states can be prepared for a given set of control Hamiltonians. This can be addressed by analyzing the dynamical Lie algebra (DLA) associated with the system~\cite{HUANG_ControllabilityQuantummechanicalSystems_1983,DAlessandro2021,Smith2025,huUniversalDynamicsGlobally2026}. If the DLA spans the full special unitary algebra, the system is controllable, and any unitary transformation can, in principle, be implemented. As a consequence, any target state can be prepared from a given initial state.

For interacting many-body systems, this analysis has been carried out only in a few cases~\cite{DAlessandro2025,huUniversalDynamicsGlobally2026}. More importantly, the DLA approach is not constructive, as it does not provide explicit pulse sequences to prepare a given target state. In addition, it yields an asymptotic notion of controllability and ignores hardware constraints such as finite control amplitudes, limited bandwidth, and finite evolution times. 

In many neutral-atom platforms, control is implemented through time-dependent external fields that couple uniformly to all qubits. These fields are limited in amplitude by hardware constraints, which restrict the dynamics and prevent arbitrarily fast qubit rotations. As a result, only a subset of Hilbert space can be explored. Understanding this restriction is essential to assess the capabilities of such simulators.

A practical way to characterize the accessible set of states is to adopt a statistical approach. States are generated using random control sequences, and the resulting ensemble statistics are compared with those of Haar-random states. We refer to this as \emph{state-expressibility}, to emphasize that we characterize the ensemble of states reachable from a fixed initial state, rather than the expressibility of the full set of unitary transformations generated by the controls. This perspective is closely related to the notion of expressibility in variational quantum circuits~\cite{Cerezo2021,Holmes2022,Sim2019}. Related statistical approaches have also been used to quantify deviations of many-body eigenstates from random-matrix predictions~\cite{Rodriguez_Nieva_2024} and to characterize the emergence of random-state statistics under unitary many-body dynamics, including Porter--Thomas statistics of measurement probabilities~\cite{Mark_2024}.

Random-state ensembles have also been observed experimentally in Rydberg systems as projected ensembles generated by chaotic many-body dynamics and used for device benchmarking~\cite{Choi2023}.

Random circuits and random-circuit sampling have been extensively studied as models for generating complex many-body dynamics~\cite{Fisher2023,NAHUM_OperatorSpreadingRandom_2018,VONKEYSERLINGK_OperatorHydrodynamicsOTOCs_2018,aruteQuantumSupremacyUsing2019}. In contrast, similar approaches for analog quantum simulators under hardware constraints remain much less explored~\cite{TANGPANITANON_ExpressibilityTrainabilityParametrized_2020,huUniversalDynamicsGlobally2026}.

In this work, we investigate how state-expressibility depends on interaction strength, evolution time, and hardware constraints. We then ask whether regimes displaying stronger signatures of state-expressibility also correspond to more favorable control landscapes, in which target states can be prepared more easily. To address this question, we use gradient-based optimization and study how the preparation difficulty depends on the properties of the target states, in particular their entanglement.

The rest of the paper is organized as follows. In Sec.~\ref{sec:random_states}, we introduce the Hamiltonian, the pulse sequences (Sec.~\ref{ssec:H}), and the hardware constraints (Sec.~\ref{ssec:hardware_constraints}) considered in this study. We then analyze the statistical properties of the generated states using entanglement entropy (Sec.~\ref{ssec:EE}), spectral statistics of reduced density matrices (Sec.~\ref{ssec:level_spacing}), and the distribution of bitstring probabilities and their relation to Porter--Thomas statistics (Sec.~\ref{ssec:PT}). The role of the Rydberg blockade is discussed in Sec.~\ref{ssec:blockade}.

In Sec.~\ref{sec:QC}, we investigate the control landscape using gradient-based optimization (GRAPE) with multiple random initializations for each target state. By selecting target states with controlled entanglement, we assess the difficulty of fixed-time state preparation under constrained control and identify regimes where highly entangled states remain hard to prepare. The optimization procedure is described in Sec.~\ref{ssec:optim}, and the results are presented in Sec.~\ref{ssec:optim_results}.

Finally, Sec.~\ref{sec:conclusion} summarizes the results and presents concluding remarks.

\section{Generation of random-pulse states}
\label{sec:random_states}

\subsection{Hamiltonian and random pulse sequences}
\label{ssec:H}

We consider an array of \(N\) Rydberg atoms arranged in a ring geometry. Each atom is treated as a two-level system (qubit), whose basis states are the ground state \(\ket{g}=\ket{\downarrow}\) and a Rydberg excited state \(\ket{r}=\ket{\uparrow}\). The system can be modeled as a collection of \(N\) spin-\(\tfrac{1}{2}\) particles interacting via a van der Waals Ising Hamiltonian~\cite{Browaeys2020,MORGADO_QuantumSimulationComputing_2021},
\begin{equation}
    \hat{H}_0 = V(d) \sum_{i<j} \frac{1}{(r_{ij}/d)^6} \hat{n}_i \hat{n}_j.
    \label{eq:H0}
\end{equation}
Here, \(d = r_{01}\) denotes the nearest-neighbor distance, and \(r_{ij}\) is the Euclidean distance between atoms at sites \(i\) and \(j\). The coupling strength is given by \(V(d) = C_6 / d^6\), where \(C_6\) is the van der Waals coefficient determined by the chosen Rydberg state (see Sec.~\ref{ssec:hardware_constraints}). The operator \(\hat{n}_i = \ket{\uparrow}_i\!\bra{\uparrow}\) is the Rydberg occupation number at site \(i\).\\

To control the collective dynamics, two external time-dependent fields (e.g., laser driving) are applied, leading to the total Hamiltonian
\begin{equation}
\label{eq:total_hamiltonian}
    \hat{H}(t) = \hat{H}_0 + \Omega(t)\hat{J}_x - \Delta(t)\hat{N}.
\end{equation}
Here, \(\hat{J}_\mu = \tfrac{1}{2} \sum_i \hat{\sigma}_\mu^{(i)}\) (\(\mu = x,y,z\)) are collective spin operators, and \(\hat{N} = \hat{J}_z + N/2 = \sum_i \hat{n}_i\) is the total Rydberg excitation number. The function \(\Omega(t) \ge 0\) is the Rabi frequency, while \(\Delta(t)\) is the detuning. Thus, the Rydberg-atom simulator is modeled by the (time-independent) atom positions (which determine the $r_{ij}$ and $d$) and two user-defined control functions $\Omega(t)$ and $\Delta(t)$. 

The control fields are modeled as piecewise-constant functions with \(M\) intervals. The time-evolution operator then factorizes as
\begin{equation}
\label{eq:unitary}
    \hat{U} = \prod_{i=1}^{M} \hat{U}_i,
\end{equation}
with ($\hbar$ set to unity):
\begin{equation}
    \hat{U}_i(\delta t_i, \Omega_i, \Delta_i) = \exp\!\left[-i\, \delta t_i \left( \hat{H}_0 + \Omega_i \hat{J}_x - \Delta_i \hat{N} \right)\right].
\end{equation}

Random pulse sequences are generated by fixing \(\delta t_i = T_{\mathrm{f}}/M\) and sampling the control amplitudes independently from uniform distributions, \(\Omega_i \sim \mathcal{U}([0,\Omega_{\max}])\) and \(\Delta_i \sim \mathcal{U}([-\Delta_{\max},\Delta_{\max}])\). The maximal values $\Omega_{\max}$ and $\Delta_{\max}$ come from hardware constraints and their values are discussed in Sec.~\ref{ssec:hardware_constraints}.

We keep the number of intervals \(M\) fixed so that the number of independently drawn control parameters remains constant, while varying the segment duration \(\delta t\) through changes in \(T_f\). 

The parameter \(M\) sets the number of random collective kicks, whereas \(\delta t\) controls the effective evolution time under the interacting Hamiltonian in each segment, and thus the amount of scrambling generated per step. 

The initial state is chosen as the product state \(\ket{\psi_0} = \ket{\downarrow}^{\otimes N}\), which is a ground state of \(\hat{H}_0\) and the natural initial state on most platforms.

We consider a ring geometry, for which the spatial symmetry group is the dihedral group \(D_N\), generated by translations and reflections. The Hamiltonian \eqref{eq:total_hamiltonian} and the initial state are invariant under this symmetry, so the time-evolved states remain confined to the symmetric subspace associated with the trivial irreducible representation of \(D_N\). For comparison, the case of an open chain, where the symmetry group contains only two elements, has been investigated in Ref.~\cite{huUniversalDynamicsGlobally2026}.

We analyze states generated by random control sequences (\textit{random-pulse states}). To make contact with experimental conditions, we first specify the model parameters and hardware constraints used throughout the analysis.
\begin{figure*}[t]
    \centering
    \includegraphics[scale=1.]{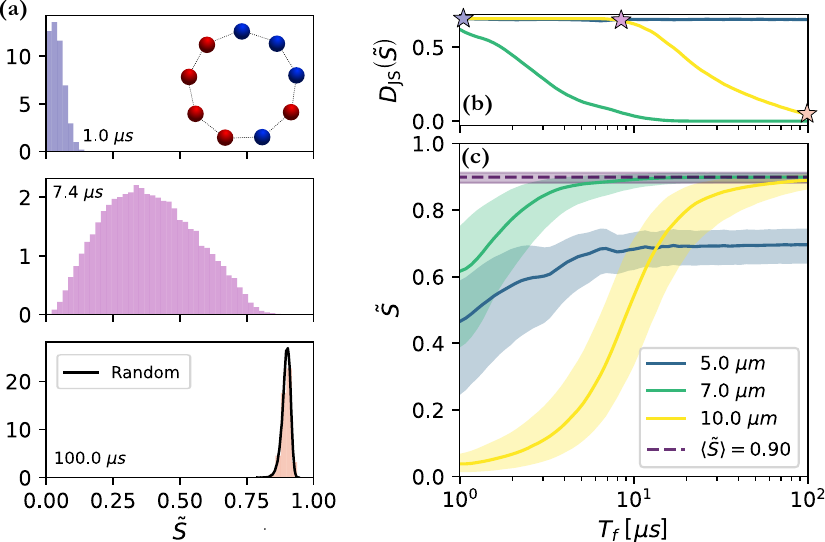}
    \caption{(a) Distributions of the normalized entanglement entropy \(\tilde{S}=S/S_{\max}\) for selected evolution times \(T_f\), compared with Haar-random states in the trivial symmetry sector. The interatomic distance is fixed to \(d=10\,\mu\mathrm{m}\). Inset: geometry of the Rydberg array, with colors indicating the bipartition used to compute \(\tilde{S}\). (b) Jensen--Shannon (JS) divergence between the distributions of \(\tilde{S}\) for random-pulse states at different \(T_f\) and those of Haar-random states in the trivial symmetry sector. (c) Median of \(\tilde{S}\) as a function of \(T_f\), with shaded regions indicating the central \(68\%\) interval. Data correspond to \(N=9\) atoms driven by piecewise-constant global pulse sequences (\(M=30\)), with amplitudes sampled as \(\Omega_k \in [0,\Omega_{\max}]\) and \(\Delta_k \in [-\Delta_{\max},\Delta_{\max}]\), where \(\Omega_{\max}=12\,\mathrm{rad}\,\mu\mathrm{s}^{-1}\) and \(\Delta_{\max}=20\,\mu\mathrm{s}^{-1}\).}
    \label{fig:EE_dist_r9_d10}
\end{figure*}

\subsection{Hardware constraints and parameters}
\label{ssec:hardware_constraints}

The control parameters are the Rabi frequency $\Omega$ and the detuning $\Delta$. Their maximum values are set by hardware constraints. Throughout this work, we use $\Omega_{\max}=12\,\mathrm{rad}\,\mu\mathrm{s}^{-1}$ and $\Delta_{\max}=20\,\mu\mathrm{s}^{-1}$, which are representative of current devices. For Pasqal platforms, see Refs.~\cite{silverioPulserOpensourcePackage2022,pulser_doc,dalyacNoiseawareEmulationCrossdevice2026}. For QuEra's platform, see Ref.~\cite{aquila_2023}.

Actually, current devices allow for larger values of $\Delta$. However, choosing $\Delta \gg \Omega$ does not significantly modify the properties of the generated states (see Sec.~\ref{ssec:blockade}). We therefore set $\Delta_{\max}$ to a value comparable to $\Omega_{\max}$.

In current platforms, atoms are trapped in optical tweezers, allowing for programmable geometries subject to a minimal spacing constraint. In our simulations, we consider a ring geometry with $N=9$ atoms and vary the nearest-neighbor distance in the range $5\,\mu\mathrm{m} \leq d \leq 10\,\mu\mathrm{m}$, consistent with experimental capabilities.

The distance $d$ sets the interaction scale $V(d)$ through the van der Waals coefficient $C_6$ (see below Eq.~\eqref{eq:H0}). Varying $d$ tunes the relative strength of interactions with respect to the control fields and thus the dynamical regime. At short distances, the system enters a blockade regime where interactions strongly constrain the dynamics (see Sec.~\ref{ssec:blockade}), while at larger distances the dynamics is dominated by the control fields.

Finally, the total evolution time is limited by the coherence time of the system, typically of order $20\,\mu\mathrm{s}$ for current Rydberg platforms (see \cite{erbinManybodyQuantumScore2026}). In this work, some of the evolution times considered exceed this scale and should therefore be interpreted as idealized dynamics.


\section{State-expressibility signatures}

We characterize state-expressibility by comparing the generated ensemble with Haar-random states within the relevant symmetry sector. We consider the entanglement entropy, the level statistics of the entanglement spectrum, and the distribution of bitstring probabilities. For these diagnostics, we also use pooled statistics over the generated states to probe the ensemble as a whole.

\subsection{Entanglement entropy}
\label{ssec:EE}

For the pure states considered here, the entanglement entropy provides a measure of bipartite entanglement and allows us to assess how close the resulting states are to Haar-random states.

Under the hardware constraints introduced above, we generate ensembles of random-pulse states for \(M=30\). We vary the total evolution time \(T_f\) and the nearest-neighbor distance \(d\), and sample \(5\times 10^4\) realizations for each parameter set. The resulting state is denoted by \(\ket{\psi(T_f)}\), and we consider a bipartition into subsystems \(A\) and \(B\) with sizes \(|A|=\lfloor N/2 \rfloor\) and \(|B|=\lceil N/2 \rceil\).

The Schmidt coefficients \(\lambda_n\) are obtained via singular value decomposition (SVD), yielding the entanglement spectrum \(s_n = -\log(\lambda_n^2)\) and the von Neumann entanglement entropy \(EE = -\sum_n \lambda_n^2 \log(\lambda_n^2)\). The bipartition (inset in Fig.~\ref{fig:EE_dist_r9_d10}(a)) is chosen such that no symmetry is induced within either subsystem. Otherwise, symmetries would lead to block structures in the reduced density matrix, affecting the level statistics (see App.~\ref{sec:ESRS}).

 For \(d=10\,\mu\mathrm{m}\), Fig.~\ref{fig:EE_dist_r9_d10}(a) shows the distribution of the normalized entanglement entropy \(\tilde{S}=S/S_{\max}\), with \(S_{\max}=N_A \log 2\). At short times (\(T_f=1.0\,\mu\mathrm{s}\)), \(\tilde{S}\) is typically small, reflecting that interactions have not yet generated significant entanglement. At intermediate times (\(T_f=7.4\,\mu\mathrm{s}\)), the distribution becomes broad. For long times (\(T_f=100\,\mu\mathrm{s}\)), it approaches that of Haar-random states within the trivial symmetry sector. The mean value is slightly reduced, from \(0.910\,S_{\max}\), expected from Page’s theorem~\cite{Hayden2006,pageAverageEntropySubsystem1993}, to \(0.897\,S_{\max}\). As discussed in Ref.~\cite{Nakata2020}, constraints on the accessible Hilbert space reduce the typical entanglement entropy. In particular, translation symmetry lowers the average value compared to the unconstrained Haar-random case.

To quantify convergence toward this distribution, we analyze the dependence on the interatomic distance (d) and total evolution time \(T_f\). We use the Jensen--Shannon (JS) divergence, a bounded and symmetric measure, to compare the entanglement entropy distribution of the random-pulse states with that of Haar-random states in the trivial symmetry sector. A related comparison of entanglement-entropy distributions using the Kullback--Leibler divergence was employed in Ref.~\cite{Rodriguez_Nieva_2024} to quantify deviations of midspectrum eigenstates from random-matrix predictions. Fig.~\ref{fig:EE_dist_r9_d10}(b) shows that for \(d=7\) and \(d=10,\mu\mathrm{m}\), the JS divergence decreases with increasing \(T_f\) and saturates at small values, indicating convergence toward Haar-like entanglement statistics at long times.

This behavior is complemented by Fig.~\ref{fig:EE_dist_r9_d10}(c), which shows the median entanglement entropy together with the central \(68\%\) interval as a function of \(T_f\). For \(d=10\,\mu\mathrm{m}\), the range of accessible entropies remains broad across timescales. At short and intermediate times (\(T_f \lesssim 10\,\mu\mathrm{s}\)), the states are weakly entangled, while at long times (\(T_f \sim 100\,\mu\mathrm{s}\)) they become highly entangled and approach Haar-like values. Around \(T_f \sim 10\,\mu\mathrm{s}\), the spread is maximal, indicating that the dynamics can generate a wide range of entanglement. This reflects a balance between interaction and control scales at \(d=10\,\mu\mathrm{m}\), leading to a diverse ensemble of weakly and highly entangled states.

\subsection{Level spacing statistics of the entanglement spectrum}
\label{ssec:level_spacing}

Pure states \(\ket{\psi}\) drawn at random from the Hilbert space, referred to as Haar-random states, exhibit a volume-law scaling of entanglement entropy~\cite{pageAverageEntropySubsystem1993}. Such behavior is also expected for states generated by sufficiently scrambling many-body dynamics~\cite{Swingle2016,gu2024simulatingquantumchaoschaos}. For these states, the reduced density matrix of a subsystem \(A\), \(\rho_A = \Tr_B\!\left(\ket{\psi}\bra{\psi}\right)\), is described by a normalized Wishart ensemble, \( \rho_A \sim X X^\dagger / \Tr(X X^\dagger) \), where \(X\) is drawn from a Ginibre ensemble~\cite{Zyczkowski2001}. Consequently, the level-spacing statistics of the entanglement spectrum follow Wigner--Dyson statistics. More precisely, the reduced density matrix belongs to the Laguerre unitary ensemble (LUE), whose local spectral statistics coincide with those of the Gaussian unitary ensemble (GUE)~\cite{Forester2010}.

The presence of level repulsion in the spectrum of \(\rho_A\) (the entanglement spectrum) reflects the random matrix character of \(\rho_A\), and thus of the state \(\ket{\psi}\). Level repulsion is therefore associated with strong entanglement, and analyzing the level statistics of \(\rho_A\) provides a way to probe the degree of scrambling in the underlying dynamics~\cite{Chamon2014,True2022}. In the following, we investigate to what extent such features can be realized in Rydberg atom arrays with global control.

\begin{figure}[H]
    \centering
    \includegraphics[scale=1.]{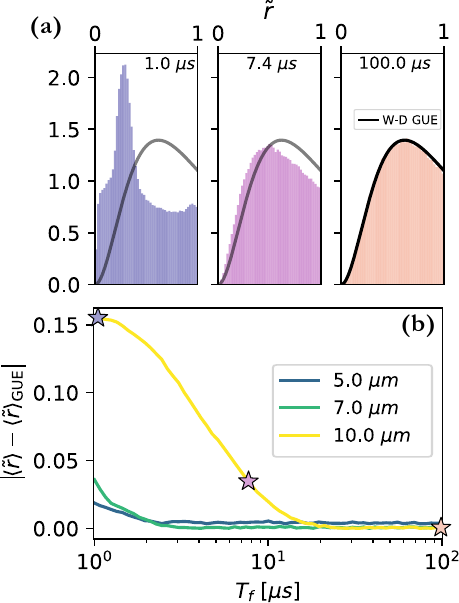}
    \caption{(a) Probability distribution of the normalized level-spacing ratio \(\tilde r\) (Eq.~\eqref{eq:rn}) for states generated by random pulse sequences, shown for three evolution times \(T_f\) at fixed interatomic distance \(d=10\,\mu\mathrm{m}\). The solid line corresponds to the Wigner--Dyson (GUE) surmise (Eq.~\eqref{eq:PWD}). (b) Absolute difference between the mean normalized level-spacing ratio \(\langle \tilde r \rangle\) and the GUE prediction, as a function of \(T_f\) for different interatomic distances \(d\). Star markers indicate the values of \(T_f\) corresponding to panel (a).}
    \label{fig:EE_ratio_r9}
\end{figure}

For each sampled state, we compute the entanglement spectrum and keep only the central \(75\%\) of the eigenvalues (the bipartition is that of Fig.~\ref{fig:EE_dist_r9_d10}). 
From this subset, we compute the normalized level-spacing ratios,
\begin{equation}
    \tilde{r}_n = \min\left( \frac{s_n - s_{n-1}}{s_{n+1} - s_n}, \frac{s_{n+1} - s_n}{s_n - s_{n-1}} \right).
    \label{eq:rn}
\end{equation}
The upper panels of Fig.~\ref{fig:EE_ratio_r9} show the pooled distribution of \(\tilde{r}\). The latter is obtained by aggregating the ratios from all states generated at fixed \(T_f\), following the procedure of Ref.~\cite{True2022}. 

To analyze the numerical data, we compare it to the Wigner--Dyson (W-D) surmise for the ratio of consecutive level spacings, given by~\cite{Atas2013}
\begin{equation}
    P_{\text{W-D}}(\tilde{r}) = \frac{2(\tilde{r} + \tilde{r}^2)^{\beta}}{Z\,(1+\tilde{r}+\tilde{r}^2)^{1+3\beta/2}},
    \label{eq:PWD}
\end{equation} 
where \(Z=4\pi/(81\sqrt{3})\) and \(\beta=2\) corresponds to GUE. As shown in Fig.~\ref{fig:EE_ratio_r9}(a), the ratio distribution for random-pulse states approaches \(P_{\text{W-D}}\) at long times. 

In addition to the evolution time \(T_f\), the interatomic distance \(d\) controls the interaction strength and thus the degree of effective constraints in the dynamics. It is therefore natural to expect that it also influences the convergence toward random-matrix behavior, as quantified in Fig.~\ref{fig:EE_ratio_r9}(b). Appendix~\ref{sec:ESRS} presents complementary data for different choices of the bipartition. In particular, we show that if the subsystem \(A\) and its complement \(B\) are symmetric with respect to each other, the level statistics is instead described by the Gaussian orthogonal ensemble (GOE).

\subsection{Bitstring probabilities and Porter--Thomas distribution}
\label{ssec:PT}

In this section, we consider the bitstring probabilities \(p(\sigma)=|\langle \psi(T_f) | \sigma \rangle|^2\), 
where \(\sigma=(\sigma_1 \cdots \sigma_N)\) denotes a computational-basis bitstring. To analyze these Born probabilities, we study the distribution \(P(\omega)\) of the rescaled variable \(\omega = D\, p(\sigma)\), where \(D=2^N\) is the Hilbert space dimension. For Haar-random states, this distribution follows the Porter--Thomas (P--T) law, \(P_{\rm PT}(\omega)=\exp(-\omega)\)~\cite{porter1956fluctuations}. The predominance of small values of \(\omega\) is known as anticoncentration~\cite{hangleiterAnticoncentrationTheoremsSchemes2018,boixo2018characterizing}. Porter--Thomas statistics of bitstring probabilities have recently been studied in a Rydberg simulator~\cite{Shaw2025}.

In random circuit sampling~\cite{hangleiterAnticoncentrationTheoremsSchemes2018,hangleiterComputationalAdvantageQuantum2023}, it is common to compare an observed distribution \(P_{\rm exp}(\omega)\) with an ideal distribution \(P_{\rm th}(\omega)\). In experiments, this comparison is often performed indirectly via cross-entropy benchmarking, which provides a quantitative measure of the agreement between measured and ideal output distributions~\cite{boixo2018characterizing,aruteQuantumSupremacyUsing2019,SAULIERE_UniversalityAnticoncentrationNoisy_2026}.

\begin{figure}[H]
    \centering
    \includegraphics{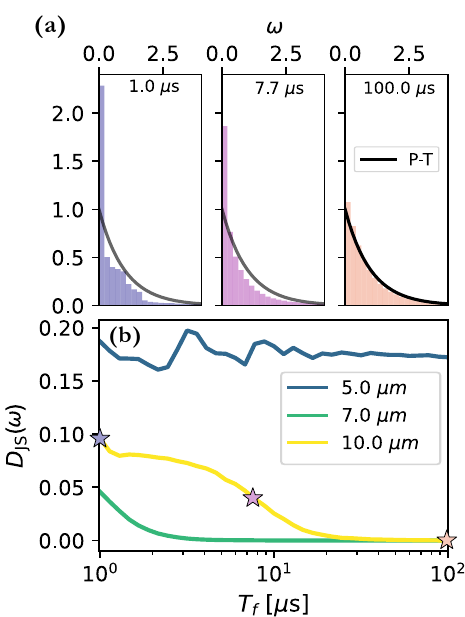}
    \caption{(a) Histograms of the normalized Born probabilities \(\omega = D\,p(\sigma)\), obtained from pooled data over computational-basis bitstrings and random-pulse realizations at selected evolution times \(T_f\), for an interatomic distance \(d=10\,\mu\mathrm{m}\), with \(D=2^N\). The solid line indicates the Porter--Thomas (P--T) distribution, \(P_{\mathrm{PT}}(\omega)=\exp(-\omega)\). (b) Jensen--Shannon (JS) divergence between the distributions of \(\omega\) for random-pulse states at different \(T_f\) and those of Haar-random states in the trivial symmetry sector, shown for different interatomic distances \(d\). Star markers indicate the values of \(T_f\) corresponding to panel (a).}
    \label{fig:pt_n9}
\end{figure}

Here, since the system size is relatively small (\(N=9\)), we can estimate \(P(\omega)\) directly by pooling data over all computational-basis bitstrings \(\sigma\) and over an ensemble of random-pulse states \(\ket{\psi(T_f)}\). Our results are summarized in Fig.~\ref{fig:pt_n9}. The top panels show histograms \(P(\omega)\) for states generated by random pulses of various durations at an interatomic distance \(d=10\,\mu\mathrm{m}\). At short and intermediate times (\(T_f=1\,\mu\mathrm{s}\) and \(7.7\,\mu\mathrm{s}\)), the distribution deviates from the P--T law, whereas at long times (\(T_f=100\,\mu\mathrm{s}\)) it agrees well with the expectation for Haar-random states.

The bottom panel quantifies this convergence for different \(d\) by displaying the JS divergence between the distributions obtained from random-pulse states at different \(T_f\) and those of Haar-random states within the trivial symmetry sector. The JS divergence between the latter and the P--T distribution is of order \(10^{-3}\), indicating that finite-size and discretization effects are negligible at the scale of the figure.

A subtle aspect of this analysis is that the generated states belong to the trivial irreducible representation of the lattice symmetry group, and therefore do not explore the full Hilbert space. Since symmetry-related bitstrings are not independent, one might expect deviations from the P--T distribution, and it is thus not obvious a priori that the Born probabilities in the \(z\)-basis should exhibit the same statistics as fully Haar-random states. Nevertheless, we find numerically that at long times the distribution \(P(\omega)\) agrees with the P--T law within numerical precision. A related analysis of bitstring probabilities in a quantum Ising model was recently reported in Ref.~\cite{huUniversalDynamicsGlobally2026} for an open chain (without translation symmetry).

\subsection{Rydberg Blockade and Constraints on State Generation}
\label{ssec:blockade}

Reducing the interatomic distance shifts the point of maximal spread of the entanglement entropy distribution to shorter evolution times, as observed for \(d=7\,\mu\mathrm{m}\), reflecting faster entanglement growth due to stronger interactions. One might therefore expect that further decreasing \(d\) would continue to shorten the timescale required to explore a broad range of entanglement values. However, this trend breaks down at \(d=5\,\mu\mathrm{m}\). As shown in Fig.~\ref{fig:EE_dist_r9_d10}(b), the JS divergence no longer decreases with increasing \(T_f\), in contrast to the behavior at \(d=7\,\mu\mathrm{m}\). Consistently, Fig.~\ref{fig:EE_dist_r9_d10}(c) shows that the entanglement entropy remains lower and its distribution does not approach Haar-like statistics within the explored timescales.

This breakdown is also reflected in the level-spacing statistics. Fig.~\ref{fig:EE_ratio_r9} shows the deviation of the mean ratio \(\langle \tilde r \rangle\) from the GUE value, \(\langle \tilde r \rangle_{\mathrm{GUE}} \approx 0.59\). For \(d=7\) and \(d=10\,\mu\mathrm{m}\), this deviation rapidly decreases with increasing \(T_f\), reaching values close to zero at times shorter than those required for the entanglement entropy distribution to approach Haar-like statistics. For \(d=5\,\mu\mathrm{m}\), the deviation remains small but finite over the explored timescales, indicating a slower approach to GUE behavior. Given the magnitude of the deviation and the associated statistical fluctuations, the data remain broadly consistent with GUE statistics.

In contrast, Fig.~\ref{fig:pt_n9} shows that the measurement probabilities reveal more pronounced deviations from Porter--Thomas statistics, providing a more stringent test of Haar-like behavior in this regime.

Taken together, these observations suggest that stronger interactions at short distances lead to slower convergence toward Haar-random behavior, possibly reflecting the presence of effective constraints in the dynamics. However, given the limited system size and statistical resolution, these deviations remain moderate and do not rule out eventual convergence at longer times.

\begin{figure}[t]
    \centering
    \includegraphics[scale=1.]{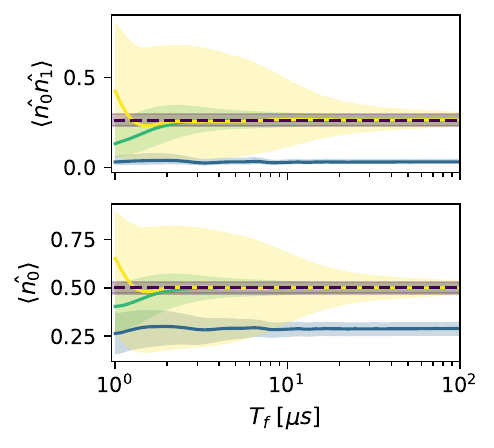}
    \caption{Median nearest-neighbor Rydberg excitation correlation $\langle \hat{n}_0 \hat{n}_1 \rangle$ obtained from quantum states generated by random pulse sequences with different total durations $T_f$. The shaded region indicates the central $68\%$ of the data.}
    \label{fig:nn_N9}    
\end{figure}

At short interatomic distances, the dynamics becomes constrained by the Rydberg blockade mechanism. We investigate how this affects the exploration of the Hilbert space by computing the nearest-neighbor Rydberg excitation correlation \(\langle \hat n_0 \hat n_1 \rangle\) for states generated by random pulse sequences, and comparing it with Haar-random states in the trivial symmetry sector. 

According to Page's theorem~\cite{pageAverageEntropySubsystem1993}, the reduced density matrix of a small subsystem of a Haar-random pure state is close to maximally mixed. For two contiguous sites, this implies \(\rho_{i,i+1} \approx \mathbb{I}_4/4\), and therefore \(\langle \hat n_i \hat n_{i+1} \rangle \approx 1/4\) and \(\langle \hat n_i \rangle \approx 1/2\). Due to translation invariance, these quantities are independent of the site index, consistent with the behavior observed in Fig.~\ref{fig:nn_N9}.

For states generated by short random pulse sequences, we observe that for large \(d\) the distributions of \(\langle \hat n_0 \hat n_1 \rangle\) and \(\langle \hat{N} \rangle / N\) are broad. Together with the low entanglement entropy in this regime, this indicates that the generated states remain weakly entangled and close to product states, with magnetization pointing in arbitrary directions, as expected for weak interactions at short times. 

As the interatomic distance is reduced to \(d=7\,\mu\mathrm{m}\), the interaction strength increases and both distributions become narrower, indicating a more constrained set of states with residual polarization inherited from the initial state. At longer evolution times, both observables approach the values expected for Haar-random states in the trivial sector.

For shorter distances, \(d=5\,\mu\mathrm{m}\), we instead observe behavior consistent with a blockade regime, where double excitations of neighboring atoms are strongly suppressed and \(\langle \hat n_0 \hat n_1 \rangle \approx 0\) over all explored timescales. 

To characterize this regime, we introduce the dimensionless parameter
\begin{equation}
    \eta = \dfrac{\Omega}{|V(d) - \Delta|},
    \label{eq:eta}
\end{equation}
which compares the driving strength \(\Omega\) to the interaction-induced detuning \(|V(d) - \Delta|\). This parameter controls the off-resonant coupling to the doubly excited state \(\ket{rr}\). In the regime \(\eta \ll 1\), the state \(\ket{rr}\) is strongly detuned and therefore weakly populated, corresponding to the Rydberg blockade regime~\cite{Urban2009}. For \(\eta \sim 1\), the blockade is partially lifted, whereas for \(V(d) \approx \Delta\), the system reaches the antiblockade condition, where \(\ket{rr}\) becomes resonantly accessible~\cite{Amthor2010,Ates2007}.

Since the Rabi frequency and detuning are drawn from uniform distributions, the probability density function of \(\eta\) can be derived analytically (see Appendix~\ref{sec:eta_prob}, Eqs.~\eqref{eq:eta_V_gt_Delta} and \eqref{eq:eta_V_le_Delta}).

\begin{figure}[t]
    \centering
    \includegraphics[scale=1.]{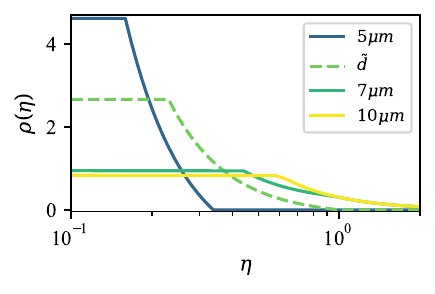}
    \caption{Probability that the drive--detuning ratio (defined in Eq.~\eqref{eq:eta}) is smaller than $\eta$, for different nearest-neighbor distances.}
    \label{fig:prob_N9}    
\end{figure}

Fig.~\ref{fig:prob_N9} shows the resulting probability density function (PDF) of \(\eta\) for the considered nearest-neighbor distances. For \(d = 5\,\mu\mathrm{m}\), the probability of sampling pulses with \(\eta \gtrsim 1\) is negligible. This is consistent with the observation that, for all evolution times, \(\mathbb{E}\langle \hat{n}_0 \hat{n}_1 \rangle \approx 0\), indicating that the system remains deep in the blockade regime.

For larger distances, the PDFs become similar, and the probability of sampling pulses with \(\eta \sim 1\) is finite, allowing for a partial lifting of the blockade.

We further identify the characteristic distance
\begin{equation}
    \label{eq:charac_d}
    \tilde{d} = \left( \frac{C_6}{\Omega_{\max} + \Delta_{\max}} \right)^{1/6},
\end{equation}
which corresponds to the shortest distance at which the sampled pulses can reach \(\eta \sim 1\),
marking the crossover between a regime in which the dynamics is always blockade-constrained and one in which the blockade can be locally lifted by the drive.

Taken together, these results reveal three main regimes of state-expressibility under random global control. At large interatomic distances, the generated ensemble progressively approaches Haar-like statistics as the evolution time increases. Reducing the distance enhances the interactions and accelerates this convergence, leading to stronger signatures of state-expressibility at shorter times. However, this trend breaks down at sufficiently short distances, where the Rydberg blockade constrains the dynamics and limits the exploration of the Hilbert space. Although the Hilbert space itself is not reduced, the finite control amplitudes restrict the region that can be explored dynamically, resulting in persistent deviations from Haar-like statistics and suppressed entanglement growth, analogous to behavior observed in constrained systems~\cite{Nakata2020}. The parameter \(\eta\) provides a quantitative characterization of the onset of this blockade-constrained regime.

\section{From state-expressibility to target-state preparation}
\label{sec:QC}

The previous section identified two regimes that are useful for studying state preparation. At \(d=10\,\mu\mathrm{m}\), varying the total evolution time at a fixed number of pulse intervals generates target states spanning a broad range of entanglement entropies. At \(d=7\,\mu\mathrm{m}\), the random-pulse ensemble approaches Haar-like statistics in the trivial symmetry sector on experimentally relevant timescales.

\begin{figure}[H]
    \centering
    \includegraphics[scale=1.]{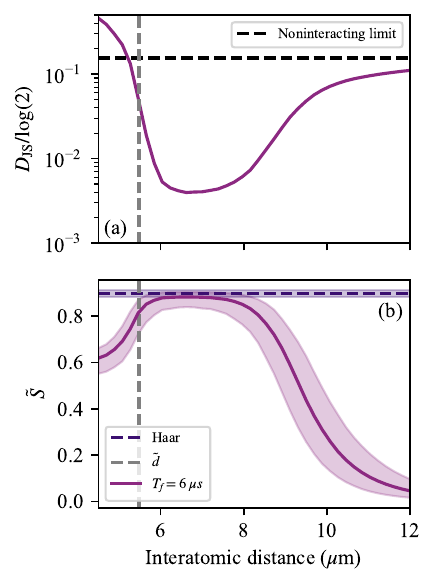}
    \caption{State-expressibility statistical diagnostics for \(T_f = 6\,\mu\mathrm{s}\) at different interatomic distances \(d\) for \(N=9\). (a) Jensen--Shannon divergence \(D_{\mathrm{JS}}/\log 2\) from the Porter--Thomas distribution. The black dashed line denotes the noninteracting, global-rotation limit. (b) Median entanglement entropy \(\tilde{S}\), with the shaded region indicating the central \(68\%\) interval. The purple dashed line denotes the Haar-random reference value, and the vertical gray dashed line marks \(\tilde{d}\) [Eq.~\eqref{eq:charac_d}].}
    \label{fig:express_fixedtime}
\end{figure}

In the preparation problem considered below, \(T_{\max}=6\,\mu\mathrm{s}\) sets the maximum time available to the optimizer, while the actual preparation time may be shorter. It is therefore useful to characterize the state-expressibility reached at this upper time scale, since this indicates whether the admissible control window extends into a regime of high state-expressibility. Figure~\ref{fig:express_fixedtime} shows that, at \(T_f=6\,\mu\mathrm{s}\), the closest approach to Haar-like behavior is obtained at intermediate distances, around \(d=7\,\mu\mathrm{m}\), while the improvement at smaller distances is limited by the onset of Rydberg blockade near the characteristic distance \(\tilde d\), defined in Eq.~\eqref{eq:charac_d}. We therefore choose \(d=7\,\mu\mathrm{m}\) as the preparation distance.

We then ask how efficiently individual target states can be prepared in this regime. Targets are drawn from the random-pulse ensemble generated at \(d=10\,\mu\mathrm{m}\), with the generation time \(T_f\) used to vary their properties, while \(T_{\max}\) sets the maximum duration available for their preparation. We analyze the resulting fidelity as a function of the target-state entanglement, the available control time, the interatomic distance, and the control-amplitude constraints.

\subsection{Optimal Control Protocol}
\label{ssec:optim}

The optimization protocol is based on \textit{GRAPE}~\cite{Khaneja2005,Rembold2020} and we follow here closely the approach of~\cite{Carrera2025}. The control pulses are taken to be piecewise constant, such that the time-evolution operator is given by Eq.~\eqref{eq:unitary}. The algorithm iteratively updates the control parameters \(\Vec{\Delta} = (\Delta_k)_{k=1}^{M}\), \(\Vec{\Omega} = (\Omega_k)_{k=1}^{M}\), and \(\Vec{\delta t} = (\delta t_k)_{k=1}^{M}\) in order to minimize a cost function defined on the final state.

The physical contribution to the cost function is
\begin{equation}
    \mathcal{C}_{\text{phys}}\left(\Vec{\Delta}, \Vec{\Omega}, \Vec{\delta t} \right)
    = -\log \left| \bra{\psi_{\text{Target}}}\hat{U}\ket{\psi_0} \right|^2,
\end{equation}
which corresponds to the negative logarithm of the fidelity between the target state \(\ket{\psi_{\text{Target}}}\) and the evolved state \(\hat{U}\ket{\psi_0}\).

To promote smooth pulse sequences and limit the total evolution time, we include a regularization term
\begin{equation}
    \begin{aligned}
    \mathcal{C}_{\text{pulse}}\left(\Vec{\Delta}, \Vec{\Omega}, \Vec{\delta t} \right) &= 
    a_1 \left( \norm{\dfrac{d\Vec{\Delta}}{dt}}^2_{\ell_2} + \norm{\dfrac{d\Vec{\Omega}}{dt}}^2_{\ell_2} \right)\\ 
    &\quad+ a_2 \exp\!\left[\left(\dfrac{\norm{\Vec{\delta t}}_{\ell_1}}{T_{\max}}\right)^{\alpha}\right].
\end{aligned}
\end{equation}
Here, \(d\Vec{\Delta}/dt\) and \(d\Vec{\Omega}/dt\) denote discrete time derivatives computed from consecutive segments (finite differences weighted by the segment durations). The first term penalizes rapid variations between intervals, while the second introduces an exponential penalty for total evolution times exceeding \(T_{\max}\).

Amplitude constraints are enforced via a quadratic hinge penalty,
\begin{equation}
\begin{aligned}
\mathcal{C}_{\text{amp}}\!\left(\vec{\Delta}, \vec{\Omega}\right)
&= a_3 \sum_{k=1}^{M} \Big[
\max\!\left(0, |\Delta_k| - \Delta_{\max}\right)^2 \\
&\qquad\qquad
+ \max\!\left(0, \Omega_k - \Omega_{\max}\right)^2
\Big].
\end{aligned}
\end{equation}

ensuring that \( |\Delta_k| \le \Delta_{\max} \) and \( 0 \le \Omega_k \le \Omega_{\max} \).

The total cost function is
\begin{equation}
    F = \mathcal{C}_{\text{phys}} + \mathcal{C}_{\text{pulse}} + \mathcal{C}_{\text{amp}}.
\end{equation}
The hyperparameters are chosen such that \(a_1 \ll 1\), reflecting a weak preference for smoothness, while \(a_2 \sim 10^{-2}\) and \(a_3 \sim 10\) strongly penalize long evolution times and amplitude violations.

\subsection{Results}
\label{ssec:optim_results}

The optimal-control procedure uses a fixed number of pulse intervals, \(M=30\). For each target state, we perform the optimization from multiple random initializations (typically 128) to reduce optimizer-dependent effects and probe the control landscape.

To define a success criterion, we first consider target states generated at the same interatomic distance used for the optimization and set \(T_{\max}=T_f\). In this case, the target is reachable by construction within the same control ansatz. We consistently obtain infidelities of order \(10^{-4}\)--\(10^{-3}\) over a broad range of target-state entanglement entropies (see App.~\ref{sec:QOC_d10}). We therefore define successful preparation using the threshold \(\gamma=2.5\times10^{-3}\).

To investigate how the difficulty of state preparation depends on the target-state entanglement, we generate 2000 target states for each \(T_f \in \{3.0,\,7.0,\,10.0,\,16.0,\,200.0\}\,\mu\mathrm{s}\) at \(d=10\,\mu\mathrm{m}\). From this pool, we perform stratified sampling over the entanglement entropy, selecting states across 30 bins to obtain an approximately uniform distribution of target entropies. The selected states are then used as targets for quantum optimal control at \(d=7\,\mu\mathrm{m}\), with multiple random initializations and a maximum control time \(T_{\max}=6\,\mu\mathrm{s}\).

\begin{figure}[H]
    \centering
    \includegraphics[scale=1.]{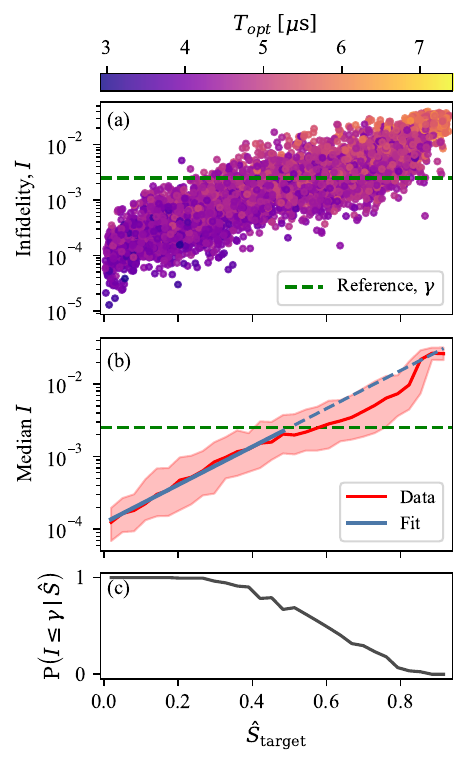}
    \caption{\textbf{(a)} Normalized entanglement entropy of the target states versus the best achieved infidelity. Target states are sampled with approximately uniform entanglement entropy (30 bins) at \(d=10\,\mu\mathrm{m}\) and generated using \(T_f \in \{3.0,\,7.0,\,10.0,\,16.0,\,200.0\}\,\mu\mathrm{s}\). The quantum optimal control algorithm is then applied at \(d=7\,\mu\mathrm{m}\) under the constraint \(T_{\max}=6\,\mu\mathrm{s}\). The color scale indicates the optimized total evolution time \(T_{\rm opt}\). \textbf{(b)} Median infidelity, with the shaded region showing the central \(68\%\) interval within each bin. \textbf{(c)} Conditional success probability \(P(I\leq\gamma\mid\hat{S}\in B_{\delta s}[\hat{S}])\), with \(\gamma=2.5\times10^{-3}\) and \(\delta s=0.0309\), where \(B_{\delta s}[\hat{S}]=[\hat{S}-\delta s/2,\,\hat{S}+\delta s/2)\).}
    \label{fig:dist_and_points_d7}
\end{figure}

Figure~\ref{fig:dist_and_points_d7}(a) shows the best achieved infidelity for each target state as a function of its normalized entanglement entropy. Although all targets reach infidelities below \(10^{-1}\), a clear dependence on the target-state entanglement is observed.

As shown in Fig.~\ref{fig:dist_and_points_d7}(b), the median infidelity increases with \(\hat{S}_{\rm target}\), with a difference of nearly two orders of magnitude between weakly entangled states, \(\hat{S}_{\rm target}\lesssim0.2\), and highly entangled states, \(\hat{S}_{\rm target}\gtrsim0.8\). At low entanglement, the increase is approximately exponential, followed by a slower variation around \(\hat{S}_{\rm target}\sim0.5\). At larger entanglement, the median infidelity rises more rapidly, while remaining below \(3\times10^{-2}\). Performing the optimization at \(d=10\,\mu\mathrm{m}\) leads to larger infidelities for highly entangled targets (see Fig.~\ref{fig:median_d10} in App.~\ref{sec:QOC_d10}).

The same trend is reflected in the conditional success probability shown in Fig.~\ref{fig:dist_and_points_d7}(c). The probability of achieving \(I\leq\gamma\) remains close to unity for weakly entangled targets, decreases gradually at intermediate entanglement, and drops sharply for highly entangled states.

These results show that, for the target-state ensemble considered here and under fixed control resources, the optimization becomes increasingly difficult as the target-state entanglement increases. Moreover, the lower infidelities obtained at \(d=7\,\mu\mathrm{m}\) compared with \(d=10\,\mu\mathrm{m}\) are consistent with the stronger state-expressibility signatures found in the previous section. As a further illustrative example, for GHZ-state preparation we find that, under the same optimization protocol, the best infidelity decreases from \(0.080\) at \(d=9\,\mu\mathrm{m}\) to \(0.012\) at \(d=7\,\mu\mathrm{m}\) (see Appendix~\ref{sec:ghz_prep}). Together, these observations suggest that operating in a regime with stronger state-expressibility signatures may facilitate the optimization of individual target states. However, the present results do not establish a causal relation between state-expressibility and optimization performance, nor do they determine whether the observed limitations originate from fundamental reachability constraints or from the finite-time control landscape.


\section{Conclusion}
\label{sec:conclusion}

We studied state-expressibility in globally controlled Rydberg atom arrays under finite-time and hardware constraints. For sufficiently long evolution times and moderate to large interatomic distances, random-pulse ensembles approach Haar-random states within the trivial symmetry sector, as indicated by the entanglement entropy, entanglement-spectrum statistics, and Born-probability distribution.

State-expressibility depends nonmonotonically on the interaction strength. Stronger interactions initially accelerate the emergence of Haar-like statistics, but at short distances the Rydberg blockade restricts Hilbert-space exploration and produces persistent deviations from Haar behavior. These deviations are more pronounced in the Born probabilities than in the entanglement-spectrum statistics, showing that the diagnostics probe complementary properties of the ensemble.

We also examined the preparation of target states drawn from random-pulse ensembles at \(d=10,\mu\mathrm{m}\). Under fixed control resources, the optimized infidelity increases with target-state entanglement. The lower infidelities obtained using dynamics at \(d=7,\mu\mathrm{m}\), compared with \(d=10,\mu\mathrm{m}\), are consistent with the stronger state-expressibility signatures in this regime, suggesting a connection between Hilbert-space exploration and target-state preparation. Overall, our results characterize how interactions, blockade constraints, and finite control resources shape state generation in Rydberg simulators beyond asymptotic controllability arguments.

Future work should determine how realistic noise sources—including positional disorder, control fluctuations, state-preparation and measurement errors, and effective Lindblad channels~\cite{dalyacNoiseawareEmulationCrossdevice2026}—affect state-expressibility and preparation performance. Extending the analysis to larger systems and different geometries would clarify the scaling and generality of the regimes identified here.


\section*{Acknowledgements}

We thank Daniel Malz for valuable comments and initial discussions, and Vincent Pasquier, Harold Erbin, Jacopo De Nardis, and Guglielmo Lami for useful discussions.
This work was granted computing time on the CCRT High-Performance Computing (HPC) facility under Grant CCRT2026-misguich awarded by the Fundamental Research Division (DRF) of CEA. This work was supported by France 2030 under Grant No.~ANR-22-PETQ-0007 (PEPR integrated project EPiQ) from the French National Research Agency (ANR).

\bibliography{thepaper}

\appendix

\appendix

\section{Entanglement spectrum ratio statistics}
\label{sec:ESRS}

In the main text, we characterized the statistical behavior of the entanglement entropy and entanglement spectrum for a fixed bipartition of the quantum system. For $N=9$ and Haar-random states in the trivial symmetry sector, the entanglement spectrum exhibits a GUE level repulsion. While the entanglement entropy of Haar-like states is largely insensitive to the geometry of the bipartition, reflecting the expected volume-law scaling in highly mixed states, we find that the level-spacing ratio statistics of the entanglement spectrum depend sensitively on this choice. Here, we analyze how the bipartition affects the gap-ratio statistics of the entanglement spectrum.

A Haar-random pure state for $N$ spin-$1/2$ systems,
\begin{equation}
    \ket{\phi} = \sum_{i_1,\dots,i_N} C_{i_1,\dots,i_N} \ket{i_1,\dots,i_N},
\end{equation}
can be constructed from coefficients drawn as independent complex Gaussian variables (Ginibre ensemble, $\beta = 2$),
\begin{equation}
    \label{eq:normal_coeff}
    \Re C_{i_1,\dots,i_N},\ \Im C_{i_1,\dots,i_N} \sim \mathcal{N}(0,1/2),
\end{equation}
followed by normalization such that $\sum_{i_1,\dots,i_N} |C_{i_1,\dots,i_N}|^2 = 1$. For a quantum system invariant under space group transformations, the tensor $C$ is fully random only within the corresponding symmetry sector, which in our case is the trivial symmetry sector.

\begin{figure}[H]
    \centering
    \includegraphics[scale=1]{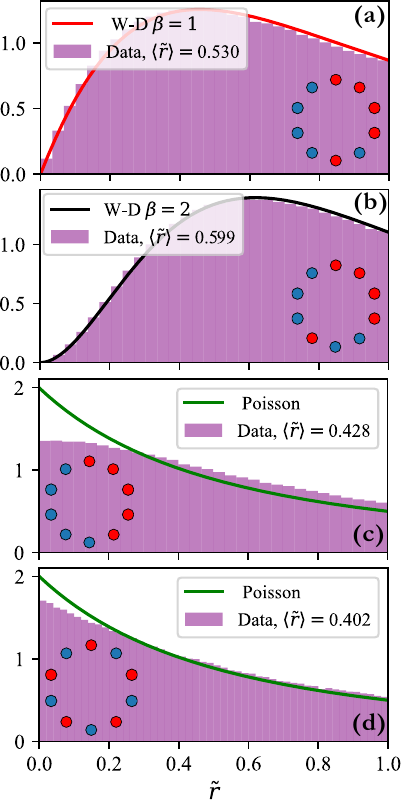}
    \caption{Distribution of the normalized level-spacing ratio $\tilde{r}$ of the entanglement spectrum, obtained from pooled data over $10^4$ Haar-random states of $N=10$ spin-$1/2$ systems in the trivial symmetric sector. Each panel corresponds to a different bipartition, illustrated in the insets. The histograms are compared to theoretical predictions for different universality classes: Wigner–Dyson distributions with $\beta=1$ (red line) and $\beta=2$ (black line), and Poisson statistics (green line). The mean value $\langle \tilde{r} \rangle$ for each dataset is indicated in the legend. The results show that the level-spacing statistics depend sensitively on the choice of bipartition, interpolating between orthogonal ($\beta=1$), unitary ($\beta=2$), and Poisson-like behavior.}
    \label{fig:ratio_10}
    
\end{figure}

We consider a bipartition of the system into subsystems $A$ and $B$ with sizes $|A|=\lfloor N/2 \rfloor$ and $|B|=\lceil N/2 \rceil$, such that
\begin{equation}
    \ket{\phi} = \sum_{i_A, i_B} C_{i_A, i_B} \ket{i_A}\otimes\ket{i_B}.
\end{equation}
Then the reduced density matrix is given by $\rho_A =\Tr_{B} \left( \ket{\phi} \! \bra{\phi} \right)$. The entanglement spectrum is obtained from the Schmidt decomposition.  

Although the global state is invariant under space group transformations, the choice of bipartition generally breaks these symmetries. Consequently, the sites assigned to subsystem $B$ influence the statistical properties of the entanglement spectrum. As shown in Fig.~\ref{fig:ratio_10}(b), when no symmetry is induced on the subsystems, the ratio statistics exhibit GUE level repulsion.

On the other hand, for bipartitions that preserve a space group transformation exchanging subsystems $A$ and $B$, the coefficient matrix satisfies
\begin{equation}
    C_{i_A,j_B}=C_{j_A,i_B},
\end{equation}
i.e., $C=C^T$. In this case, $C$ admits an Autonne--Takagi decomposition,
\begin{equation}
    C = Q \Lambda Q^{T},\qquad Q\in U(D),
\end{equation}
with $\Lambda=\mathrm{diag}(\lambda_1,\dots,\lambda_D)\geq0$ diagonal~\cite{HORN_MatrixAnalysis_2012,guo2025eigensingularcosinesineautonnetakagi}. Defining
\begin{equation}
    A = Q^\dagger dQ,
\end{equation}
unitarity of $Q$ implies
\begin{equation}
    A^\dagger = -A, \qquad A^T = -A^*.
\end{equation}
Differentiating the Takagi decomposition,
\begin{equation}
    dC = dQ \Lambda Q^T + Q d\Lambda Q^T + Q\Lambda dQ^T,
\end{equation}
and multiplying by $Q^\dagger$ from the left and $Q^*$ from the right gives
\begin{equation}
    Q^\dagger dC Q^* = A\Lambda + d\Lambda - \Lambda A^*.
\end{equation}
Using the Euclidean metric induced by the Frobenius inner product,
\begin{equation}
    ds^2=\Tr(dC\,dC^\dagger),
\end{equation}
together with cyclicity of the trace,
\begin{equation}
    ds^2
    =
    \Tr\!\left[
        (A\Lambda+d\Lambda-\Lambda A^*)
        (A^\dagger\Lambda+d\Lambda-\Lambda A)
    \right].
\end{equation}
Since $d\Lambda$ is diagonal while $A\Lambda-\Lambda A^*$ is off-diagonal, the mixed terms vanish, yielding
\begin{equation}
    ds^2
    =
    \Tr\!\left[
        (A\Lambda-\Lambda A^*)
        (A\Lambda-\Lambda A^*)^\dagger
    \right]
    +
    \Tr(d\Lambda^2).
\end{equation}
For the matrix elements,
\begin{equation}
    (A\Lambda-\Lambda A^*)_{ij}
    =
    A_{ij}\lambda_j-\lambda_i A_{ij}^*.
\end{equation}
Writing
\begin{equation}
    A_{ij}=x_{ij}+iy_{ij},
\end{equation}
with $x_{ij},y_{ij}\in\mathbb{R}$, one obtains
\begin{equation}
    A_{ij}\lambda_j-\lambda_i A_{ij}^*
    =
    (\lambda_j-\lambda_i)x_{ij}
    +
    i(\lambda_j+\lambda_i)y_{ij}.
\end{equation}
Therefore,
\begin{equation}
    \left|
    A_{ij}\lambda_j-\lambda_i A_{ij}^*
    \right|^2
    =
    (\lambda_j-\lambda_i)^2 x_{ij}^2
    +
    (\lambda_j+\lambda_i)^2 y_{ij}^2.
\end{equation}
The metric then becomes
\begin{equation}
\begin{aligned}
    ds^2
    =&
    \sum_{i<j}
    \left[
        (\lambda_j-\lambda_i)^2 dx_{ij}^2
        +
        (\lambda_j+\lambda_i)^2 dy_{ij}^2
    \right] \\
    &+
    \sum_i 4\lambda_i^2 dy_i^2
    +
    \sum_i d\lambda_i^2 .
\end{aligned}
\end{equation}
The corresponding Jacobian is therefore
\begin{equation}
    \label{eq:goe_taka}
    \sqrt{\det g}
    \propto
    \prod_{i<j} |\lambda_j^2-\lambda_i^2|
    \prod_i |\lambda_i|.
\end{equation}

Since the reduced density matrix reads
\begin{equation}
    \rho_A = C C^\dagger = Q \Lambda^2 Q^\dagger,
    \label{eq:takagi}
\end{equation}
the eigenvalues of $\rho_A$ are given by $p_i=\lambda_i^2$. Equation~\eqref{eq:goe_taka} therefore induces the probability measure
\begin{equation}
    d\mu(\rho_A)
    \propto
    \prod_{i<j}|p_i-p_j|
    \prod_i dp_i,
\end{equation}
which exhibits the characteristic linear repulsion of the orthogonal universality class ($\beta=1$). This is consistent with our numerical results shown in Fig.~\ref{fig:ratio_10}(b), where the gap-ratio statistics are well described by Eq.~\eqref{eq:PWD} with $\beta=1$. To the best of our knowledge, no previous study has reported such level statistics arising from a spatial symmetry exchanging the two subsystems of the bipartition.

We then consider bipartitions that induce spatial symmetries within the subsystems. As illustrated in Fig.~\ref{fig:ratio_10}(c,d), these symmetries lead to a block-diagonal structure of the reduced density matrix, with each block corresponding to a distinct symmetry sector. Consequently, the entanglement spectrum decomposes into independent symmetry-resolved contributions, and the resulting level-spacing statistics no longer follow a single random-matrix ensemble, but instead reflect a superposition of spectra from different symmetry sectors.

This explains the choice of bipartition used in the main text, as it isolates the desired symmetry class of the entanglement spectrum.

\section{Quantum optimal control at large distance}
\label{sec:QOC_d10}

To establish a meaningful success criterion for quantum state preparation using optimal control, we consider target states that are in principle reachable, i.e., states within the accessible region of Hilbert space. This is achieved by setting $T_{\max} = T_f$. This also provides a benchmark of the optimisation algorithm.

For each target, we apply the optimization procedure described in the main text (Sec.~\ref{ssec:optim}) to determine pulse sequences that approximate the state. We consider targets generated at $T_f = 6\,\mu\mathrm{s}$ and perform the optimization under the constraint $T_{\max} = 6\,\mu\mathrm{s}$, both at a fixed interatomic distance $d = 10\,\mu\mathrm{m}$.

\begin{figure}[H]
    \centering
    \includegraphics[scale=1.]{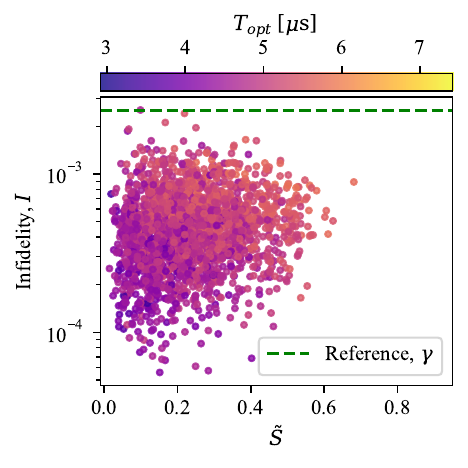}
    \caption{Normalized entanglement entropy of the target states versus the best achieved infidelity. Target states are generated at $d=10\,\mu\mathrm{m}$ with $T_f = 6\,\mu\mathrm{s}$, and the quantum optimal control algorithm is run under the same interatomic distance with $T_{\max} = T_f$.  The color scale indicates the optimal total evolution time found by the algorithm.}
    \label{fig:reference}
\end{figure}

Figure~\ref{fig:reference} shows the normalized entanglement entropy $\tilde S$ of the target states and the best achieved infidelity. We observe that low infidelities are consistently reached, with typical values of order $10^{-4}$ to $10^{-3}$. In addition, the optimized evolution times vary across instances, indicating that many target states can be prepared using a sequence that is {\em shorter} than the one used for their generation. This confirms that the optimization algorithm is efficient within the accessible region of Hilbert space. We therefore define successful preparation using a fixed infidelity reference $\gamma = 2.5 \times 10^{-3}$, chosen close to the highest observed infidelity.

\begin{figure}[H]
    \centering
    \includegraphics[scale=1.]{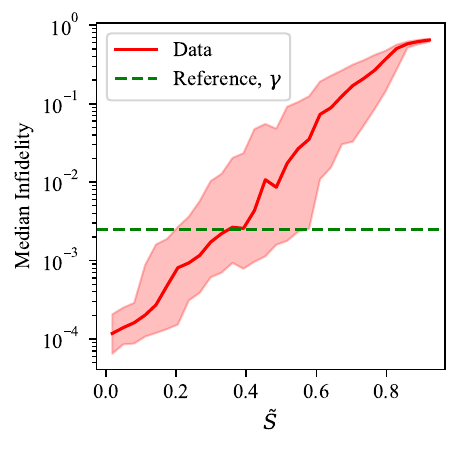}
    \caption{Same as Fig. \ref{fig:dist_and_points_d7}(b) but the optimal quantum control algorithm is run with $d = 10\,\mu\mathrm{m}$.}
    \label{fig:median_d10}
\end{figure}

Further, as discussed in the main text, the interatomic distance, together with the hardware constraints, determines an optimal value of $d$ for state preparation. For $d = 7\,\mu\mathrm{m}$, the infidelity remains low, $I \lesssim 10^{-2}$, across states with a wide range of entanglement, including highly entangled ones. In contrast, Fig.~\ref{fig:median_d10} shows that for $d = 10\,\mu\mathrm{m}$, highly entangled states, $\tilde{S} \gtrsim 0.8$ can only be poorly approached (large infidelities, $I \sim 1$).

These results show that the interatomic distance strongly affects the set of states that can be reliably prepared.

\section{Chance of getting blockade}
\label{sec:eta_prob}

We assume that the control amplitudes are drawn independently as
\begin{equation}
    \Omega \sim \mathcal U([0,\Omega_{\max}]),
    \qquad
    \Delta \sim \mathcal U([-\Delta_{\max},\Delta_{\max}]),
\end{equation}
and define
\begin{equation}
    \eta=\frac{\Omega}{|V(d)-\Delta|}.
\end{equation}
For brevity, let
\begin{equation}
    V \equiv V(d), \qquad Y \equiv |V-\Delta|,
\end{equation}
so that \(\eta=\Omega/Y\). Since \(\Omega\) and \(\Delta\) are independent, \(\Omega\) and \(Y\) are also independent.

The PDF of \(\Omega\) is
\begin{equation}
    f_{\Omega}(\omega)=\frac{1}{\Omega_{\max}},
    \qquad 0\le \omega \le \Omega_{\max}.
\end{equation}

We now derive the PDF of \(Y\), and then that of \(\eta\).

\subsection*{Distribution of \(Y=|V-\Delta|\)}

\textbf{Case A: \(V>\Delta_{\max}\).}\\
In this regime, \(V-\Delta>0\) for all \(\Delta\in[-\Delta_{\max},\Delta_{\max}]\), so
\begin{equation}
    Y=V-\Delta.
\end{equation}
Since \(\Delta\) is uniform, \(Y\) is also uniform:
\begin{equation}
    f_Y(y)=\frac{1}{2\Delta_{\max}},
    \qquad
    V-\Delta_{\max}\le y\le V+\Delta_{\max}.
\end{equation}

\textbf{Case B: \(0<V\le \Delta_{\max}\).}\\
In this case, the map \(\Delta\mapsto Y=|V-\Delta|\) is two-to-one for small \(y\), and one-to-one near the upper edge. Explicitly,
\begin{equation}
    f_Y(y)=
    \begin{cases}
        \dfrac{1}{\Delta_{\max}},
        & 0\le y\le \Delta_{\max}-V, \\[0.6em]
        \dfrac{1}{2\Delta_{\max}},
        & \Delta_{\max}-V< y\le \Delta_{\max}+V, \\[0.6em]
        0,
        & \text{otherwise}.
    \end{cases}
\end{equation}

\subsection*{Distribution of \(\eta=\Omega/Y\)}

Using the change of variables
\begin{equation}
    \Omega=\eta y,
\end{equation}
the Jacobian is
\begin{equation}
    \left|\frac{\partial(\Omega,Y)}{\partial(\eta,y)}\right|=y.
\end{equation}
Hence
\begin{equation}
    \rho(\eta)=f_{\eta}(\eta)
    =\int_0^\infty f_{\Omega}(\eta y)\,f_Y(y)\,y\,dy.
\end{equation}
Since \(f_{\Omega}(\eta y)\neq 0\) only if \(0\le \eta y\le \Omega_{\max}\), this becomes
\begin{equation}
    \rho(\eta)
    =\frac{1}{\Omega_{\max}}
    \int_0^{\Omega_{\max}/\eta} y\,f_Y(y)\,dy.
    \label{eq:eta_master}
\end{equation}

We now evaluate this integral in the two regimes above.\\

\textbf{Case A: \(V>\Delta_{\max}\).}\\

Using \(f_Y(y)=1/(2\Delta_{\max})\) on
\([V-\Delta_{\max},\,V+\Delta_{\max}]\), Eq.~\eqref{eq:eta_master} gives
\begin{equation}
    \rho(\eta)
    =\frac{1}{2\Delta_{\max}\Omega_{\max}}
    \int_{V-\Delta_{\max}}^{\min(V+\Delta_{\max},\,\Omega_{\max}/\eta)} y\,dy.
\end{equation}
It is convenient to define the thresholds
\begin{equation}
    \eta_- \equiv \frac{\Omega_{\max}}{V+\Delta_{\max}},
    \qquad
    \eta_+ \equiv \frac{\Omega_{\max}}{V-\Delta_{\max}}.
\end{equation}
Then the three regions are:

\begin{itemize}
    \item If \(0\le \eta\le \eta_-\), the whole support contributes, and
    \begin{equation}
        \rho(\eta)=\frac{V}{\Omega_{\max}}.
    \end{equation}

    \item If \(\eta_-<\eta\le \eta_+\), the upper limit is \(\Omega_{\max}/\eta\), and
    \begin{equation}
        \rho(\eta)=\frac{\Omega_{\max}}{4\Delta_{\max}}
        \left[
        \frac{1}{\eta^2}
        -\left(\frac{V-\Delta_{\max}}{\Omega_{\max}}\right)^2
        \right].
    \end{equation}

    \item If \(\eta>\eta_+\), the integration domain is empty, and
    \begin{equation}
        \rho(\eta)=0.
    \end{equation}
\end{itemize}

Therefore, for \(V(d)>\Delta_{\max}\),
\begin{equation}
\label{eq:eta_V_gt_Delta}
\rho(\eta)=
\begin{cases}
\dfrac{V(d)}{\Omega_{\max}}, & 0\le \eta \le \eta_-, \\[0.6em]

\dfrac{\Omega_{\max}}{4\Delta_{\max}}
\left[
\dfrac{1}{\eta^2}
-\left(\dfrac{V(d)-\Delta_{\max}}{\Omega_{\max}}\right)^2
\right],
& \eta_-<\eta\le \eta_+, \\[0.6em]

0, & \eta>\eta_+.
\end{cases}
\end{equation}

\textbf{Case B: \(0<V\le \Delta_{\max}\).}\\

Using the piecewise form of \(f_Y(y)\), Eq.~\eqref{eq:eta_master} becomes
\begin{align}
\rho(\eta)
&= \frac{1}{\Omega_{\max}}
\Bigg[
\int_0^{\min\{\Delta_{\max}-V,\Omega_{\max}/\eta\}}
\frac{y}{\Delta_{\max}}\,dy \\
&\quad +
\int_{\Delta_{\max}-V}^{\min\{\Delta_{\max}+V,\Omega_{\max}/\eta\}}
\frac{y}{2\Delta_{\max}}\,dy
\Bigg].
\end{align}
In this regime we define
\begin{equation}
    \eta_- \equiv \frac{\Omega_{\max}}{\Delta_{\max}+V},
    \qquad
    \eta_+ \equiv \frac{\Omega_{\max}}{\Delta_{\max}-V}.
\end{equation}
Then:

\begin{itemize}
    \item If \(0\le \eta\le \eta_-\), the whole support contributes:
    \begin{equation}
        \rho(\eta)
        =\frac{\Delta_{\max}^2+V^2}{2\Omega_{\max}\Delta_{\max}}.
    \end{equation}

    \item If \(\eta_-<\eta\le \eta_+\), the upper limit lies in the second branch:
    \begin{equation}
        \rho(\eta)
        =\frac{\Omega_{\max}}{4\Delta_{\max}}
        \left[
        \frac{1}{\eta^2}
        +\left(\frac{\Delta_{\max}-V}{\Omega_{\max}}\right)^2
        \right].
    \end{equation}

    \item If \(\eta>\eta_+\), only the first branch contributes:
    \begin{equation}
        \rho(\eta)
        =\frac{\Omega_{\max}}{2\Delta_{\max}}\frac{1}{\eta^2}.
    \end{equation}
\end{itemize}

Thus, for \(V(d)\le \Delta_{\max}\),
\begin{equation}
\label{eq:eta_V_le_Delta}
\rho(\eta)=
\begin{cases}
\dfrac{\Delta_{\max}^{2}+V(d)^{2}}{2\,\Omega_{\max}\Delta_{\max}},
& 0\le \eta\le \eta_-, \\[0.6em]

\dfrac{\Omega_{\max}}{4\,\Delta_{\max}}
\left[
\dfrac{1}{\eta^2}
+\left(\dfrac{\Delta_{\max}-V(d)}{\Omega_{\max}}\right)^2
\right],
& \eta_-<\eta\le \eta_+, \\[0.6em]

\dfrac{\Omega_{\max}}{2\,\Delta_{\max}}\dfrac{1}{\eta^{2}},
& \eta>\eta_+.
\end{cases}
\end{equation}

In the limiting case \(V(d)=\Delta_{\max}\), one has \(\eta_+\to\infty\), so the third branch disappears.
\section{GHZ-state preparation}
\label{sec:ghz_prep}

\begin{figure}[H]
    \centering
    \includegraphics[scale=1.]{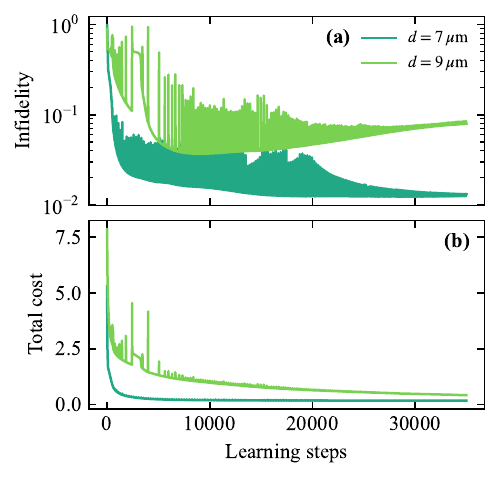}
    \caption{
    Learning curves for GHZ-state preparation at interatomic distances
    \(d=7\,\mu\mathrm{m}\) and \(d=9\,\mu\mathrm{m}\) for \(N=9\) and
    \(T_{\max}=6\,\mu\mathrm{s}\). Both optimizations use the NadamW optimizer
    with learning rate \(2.5\times10^{-3}\), the default Optax hyperparameters
    otherwise, and the same number of learning steps.
    \textbf{(a)} Infidelity as a function of the learning step. The best
    infidelities obtained are \(0.012\) and approximately \(0.080\) for
    \(d=7\,\mu\mathrm{m}\) and \(d=9\,\mu\mathrm{m}\), respectively.
    \textbf{(b)} Corresponding total cost during the optimization.
    }
    \label{fig:ghz}
\end{figure}

In Sec.~\ref{ssec:optim_results}, we investigated whether the
state-expressibility diagnostics can help identify interaction regimes that are
favorable for constrained state preparation. That analysis was performed using
target states generated by random pulse sequences at \(d=10\,\mu\mathrm{m}\).
Here, we complement those results with an illustrative example involving a
structured many-body target state, namely the GHZ state.

For the bipartition considered here, the GHZ state has normalized entanglement entropy
\(\tilde{S}=S/S_{\max}=1/N_A=0.25\). We optimize its preparation at
\(d=7\,\mu\mathrm{m}\) and \(d=9\,\mu\mathrm{m}\), using the same optimization
protocol and a maximum control time \(T_{\max}=6\,\mu\mathrm{s}\).

Figure~\ref{fig:ghz} shows the corresponding learning curves. At
\(d=7\,\mu\mathrm{m}\), the optimization reaches a best infidelity
\(I=0.012\), whereas at \(d=9\,\mu\mathrm{m}\) the best infidelity is
\(I\simeq0.080\). Thus, for this particular target state and under the same
optimization conditions, the lower-distance regime leads to a substantially
better preparation fidelity. This observation is consistent with the stronger
state-expressibility signatures found at \(d=7\,\mu\mathrm{m}\), and provides
an example suggesting that operating in a more expressive regime may facilitate
constrained state preparation.

The best infidelity obtained at \(d=7\,\mu\mathrm{m}\) is also compatible with
the range observed in Fig.~\ref{fig:dist_and_points_d7} for random-pulse target
states with \(\tilde{S}\simeq0.25\). Nevertheless, this comparison should be
interpreted only as an illustrative example: states with the same entanglement
entropy can have very different structures and preparation difficulties, and
the present result does not establish a causal relation between
state-expressibility and optimization performance.

\end{document}